\documentclass{bmvc2k}

\usepackage{xcolor}

\definecolor{g}{RGB}{47,79,79}
\definecolor{g}{RGB}{0,0,109}

\usepackage{graphicx}
\usepackage{comment}
\usepackage{amsthm}
\usepackage{amsmath} 
\usepackage{color}
\usepackage{wrapfig}
\usepackage[utf8]{inputenc} 
\usepackage[T1]{fontenc}    
\usepackage{url}            
\usepackage{booktabs}       
\usepackage{amsfonts}       
\usepackage{nicefrac}       
\usepackage{microtype}      
\usepackage{appendix}
\usepackage{diagbox}
\usepackage{algorithm}
\usepackage[noend]{algpseudocode}
\usepackage{ragged2e,array,booktabs}
\usepackage{commath}
\usepackage{epsfig}
\usepackage{graphicx}
\usepackage{comment}
\usepackage{booktabs}       
\usepackage{microtype}
\usepackage{graphicx}
\usepackage{bm}
\usepackage{multirow}

\newcommand{\R}{\mathbb{R}}

\newcommand{\etal}{\textit{et al}. }

\title{Efficient Cross-Modal Retrieval\\ via Deep Binary Hashing and Quantization}

\addauthor{Yang Shi}{yang.shi@rakuten.com}{1}
\addauthor{Young-joo Chung}{youngjoo.chung@rakuten.com}{1}

\addinstitution{
 Rakuten, Inc.\\
 San Mateo, USA
}

\runninghead{SHI AND CHUNG}{Efficient Cross-Modal Retrieval  via HQ}

\def\etal{\emph{et al}\bmvaOneDot}

\begin{document}

\maketitle

\begin{abstract}
Cross-modal retrieval aims to search for data with similar semantic meanings across different content modalities.
However, cross-modal retrieval requires huge amounts of storage and retrieval time since it needs to process data in multiple modalities.
Existing works focused on learning single-source compact features such as binary hash codes that preserve similarities between different modalities.
  In this work, we propose a jointly learned deep hashing and quantization network (HQ) for cross-modal retrieval. We \emph{simultaneously} learn binary hash codes and  quantization codes to preserve semantic information in multiple modalities by an end-to-end deep learning architecture. At the retrieval step, binary hashing is used to retrieve a subset of items from the search space, then quantization is used to re-rank the retrieved items. We  theoretically and empirically show that this two-stage retrieval approach provides faster retrieval results while preserving accuracy. Experimental results on the NUS-WIDE, MIR-Flickr, and Amazon datasets demonstrate that HQ achieves boosts of more than 7\% in precision compared to supervised neural network-based compact coding models.
\end{abstract}

\section{Introduction}

Cross-modal retrieval aims to search for data with similar semantic meanings across different content modalities, such as audio-text tag search in music recommender systems~\cite{music-retrieval} and image-text search in web search~\cite{nuswide,figexplorer}.
In cross-modal retrieval scenarios, features from different modalities are mapped to the shared space to be compared using searching and ranking algorithms. 
However, storing and comparing features from multiple modalities requires huge computational resources because of the size of the dataset and the number of modalities~\cite{SDML-CMR}. To overcome these challenges, approximate nearest neighbor search (ANN)~\cite{ann} across different content modalities has gained huge attention~\cite{cmr-1}. ANN search based on compact coding methods, including binary hashing and quantization, is known to achieve an order of magnitude speed-ups compared to exact Nearest Neighbor (NN) search while preserving near-optimal accuracy~\cite{ann,QBH}. 


Compact coding methods learn efficient codes (usually binary values) that can substitute original continuous features. Binary hashing aims to learn binary codes that provide competitive accuracy as well as continuous features while reducing the space and the retrieval time. Following the success in single-modal retrieval~\cite{semi,iq,sdh,kernel,LIR-SBP}, methods for creating binary hash codes that can connect multiple modalities have been proposed~\cite{cmhh,scm,cmfh,svhn,ssah,HiCHNet,UKD}. Furthermore, to overcome the limitation of the expressive power of binary hashing, quantization-based methods have been explored~\cite{cdq,shared,djsrh}. Quantization learns a shared lookup table (a dictionary) consisting of continuous values as well as binary codes for each data point to indicate which dictionary entry the data point is hashed to. As a result, it obtains more accurate approximations compared to simple binary hash codes. However, such quantization-based methods sacrifice computational efficiency compared to binary hashing.



\begin{figure*}
\centering
\subfigure[Training process]{
\includegraphics[height = 3cm]{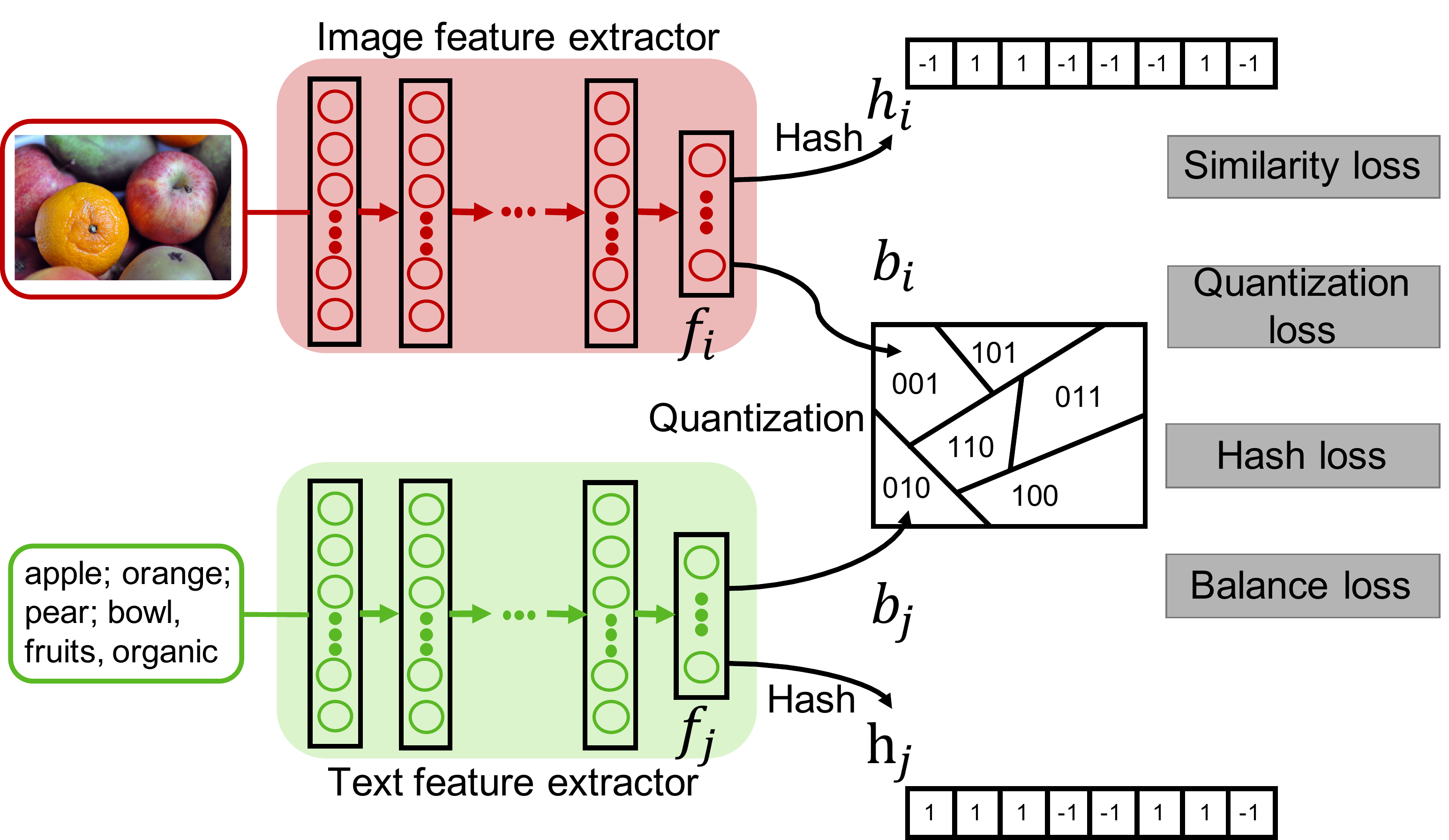}%
\label{fig:model-t}
}
\subfigure[Retrieval process]{
 \includegraphics[height = 3cm]{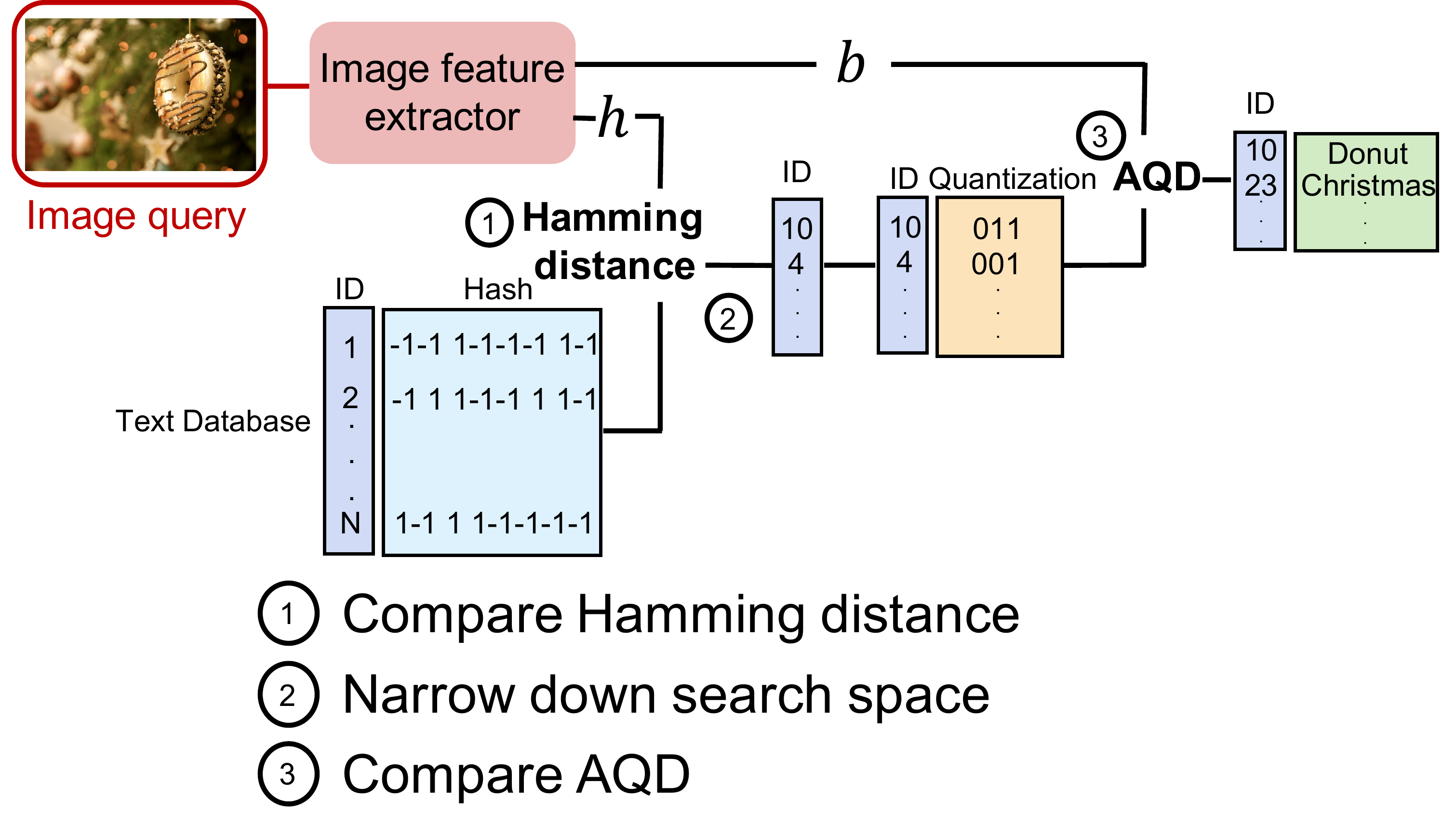}%
    \label{fig:model-r}%
}
\caption{Deep cross-modal binary Hashing and Quantization network (HQ). In the training step, we jointly learn binary codes and quantization codes for data from different modalities. In the retrieval step, we first narrow down the search space by comparing the Hamming distance of binary codes, then measure the similarity between the query and the data in the smaller search space using quantization distance.}%
\label{fig:model}%
\end{figure*}

In this paper, we propose a new supervised cross-modal retrieval model via deep binary \textbf{H}ashing and \textbf{Q}uantization (\textbf{HQ}). It combines binary hashing and quantization to fully explore each other's strengths via simultaneously learning an end-to-end deep learning network. To fully utilize both codes, we also leverage a two-stage retrieval framework that is widely used in practice~\cite{two-stage,cascade,youtube}. Thus, HQ can
provide more accurate results than binary hashing-based methods and require fewer computational resources than quantization-based methods.  
The overview of HQ is shown in Figure~\ref{fig:model}. 

Our contributions are as follows:
\begin{itemize}
\item We propose a cross-modal retrieval model via deep binary hashing and quantization (HQ). It \emph{simultaneously} learns binary hash codes and quantization codes using an end-to-end deep learning architecture to preserve semantic information in multiple modalities. Both codes are fully utilized by a two-stage retrieval, which first narrows down search space by comparing the Hamming distance of binary codes, and then measures the similarity between the query and the data in the smaller search space using quantization distance measurement.  
 \item  
    We analyze the computation and memory complexity of HQ retrieval process. We show $O(n)$ times speed-up using HQ compared to the quantization method while maintaining the same level of memory usage and retrieval accuracy ($n$ is the feature length before quantization or binary hash).

    \item We demonstrate the retrieval accuracy of HQ through extensive experiments on the NUS-WIDE, MIR-Flickr, and Amazon datasets. HQ achieved boosts of more than 7\% in precision compared to supervised neural network-based models that utilize either binary hashing or quantization. 
    
\end{itemize}

\section{Preliminaries}
\label{sec:prelim}

\textbf{Binary hashing} Given input $x \in \R^{n}$, we compute binary code $h_x = H(x) \in \R^{n}$, where $H(\cdot)$ is a function that maps continuous value to $\{+1,-1\}$. The similarity measurement is \textit{Hamming distance}:
$
  \text{dist}(h_x,h_y) = \text{sum } (h_x \text{ XOR } h_y)  
$.
 The larger the value, the greater the dissimilarity between the two.\\
\textbf{Quantization} Given input $x \in \R^{n}$, we compute the quantization $ x \approx \sum_{l=1}^mC^lb_x^l $, where $C^l \in \R^{n \times k}$ is the \textit{l-th} dictionary book, $b^l_x \in \R^{k} $ is the \textit{l-th} dictionary index indicator for $x$. We assume the input is assigned to only  one entry of the dictionary: $\norm{b^l_x}_0 = 1$. Different inputs will share the same $C^l$ but will have different index indicators.
We use \textit{Asymmetric Quantizer Distance (AQD)} to measure quantization similarities: 
$
AQD(x,y) = x^T(\sum_{l=1}^m C^lb^l_y)  $. A greater value correlates with a greater similarity between the two.
Generally, quantization preserves more information than binary codes, though it is slower at searching step since AQD requires more computation than Hamming distance.

\section{Related Work}
\subsection{Cross-modal Hashing}
With the increase of multi-modal data, the need for cross-modal retrieval increased, and creating compact codes for cross-modal retrieval has been extensively explored to provide efficient retrieval. Hashing methods such as Cross Modality Similarity Sensitive Hashing (CMSSH) ~\cite{cmr-1}, Semantic Correlation Maximization (SCM)~\cite{scm}, and Semantics Preserving Hashing (SePH)~\cite{seph}  mapped data into a common Hamming space and compared their similarities using Hamming distance. Probabilistic frameworks were also introduced to learn both the binary codes and their dimensions~\cite{ozdemir2014,Gholami2016}. 
Recently, 
hashing methods utilizing deep learning to create features have been explored. Starting with deep visual-semantic hashing~\cite{dvs} that utilized Convolutional Neural Network and Recurrent Neural Network to learn text-image modal-shared features to be hashed, Correlation Hashing Network (CHN)~\cite{p2} jointly learned data representation from each modality and used a structured max-margin loss to learn similarity-preserving and high-quality hash codes. 
Deep Cross-modal hashing (DCMH)~\cite{dcmh} utilized two deep neural networks, one for each modality, and learned the discrete hash codes directly through a sign function. 
Deep Joint-Semantics Reconstructing Hashing (DJSRH)~\cite{djsrh} introduced a joint-semantics affinity matrix to combine the semantic similarity in each modality and improve cross-modal coding similarity. Joint-modal Distribution-based Similarity Hashing (JDSH)~\cite{jdsh} further simplified DJSRH and proposed a sampling and weighting scheme to strengthen the discriminative ability in hash codes. 
UGACH~\cite{ugach}, UCH~\cite{uch} and  DADH~\cite{cong2020} used generative adversarial networks (GAN) to learn better representations from different modalities.


\subsection{Cross-modal Quantization}
To solve the limitation of the expressive power of binary hashing, quantization has been introduced for single/multi-modal retrieval~\cite{itq,cq}. Quantized correlation hashing (QCH)~\cite{qch} is the first attempt to integrate hash function learning and quantization for cross-modal retrieval. 
Composite Correlation Quantization (CCQ)~\cite{ccq}  mapped both paired images and texts to shared latent space and learned composite quantizers that convert the shared latent features into compact binary codes while preserving both intra-modal similarity and inter-modal correlation. 
Collective Deep Quantization (CDQ)~\cite{cdq} introduced a deep neural network to quantization for  cross-modal retrieval. Shared Predictive Deep Quantization (SPDQ)~\cite{shared} created a shared subspace across different modalities and private subspaces for individual modalities. Representations are learned from both subspaces.

Little research has been conducted in combining binary hashing and quantization. Quantization-based hashing (QBH)~\cite{QBH} is the first attempt to incorporate the quantization-based method into the similarity-preserve hashing. It learned hashing and quantization codes  \emph{sequentially}. 
It minimized a similarity-preserving error to learn binary hash codes. The quantization error was calculated based on the assumption that data with the same hash code should share the same dictionary. Thus, the quantization was restricted by hashing. This approach was applied to single-modal retrieval based on hand-crafted features. 

To the best of our knowledge, our work is the first attempt to integrate binary hashing and quantization in end-to-end feature learning for efficient cross-modal retrieval. Unlike previous methods, HQ learns binary hash codes and quantization codes simultaneously. We theoretically and empirically demonstrate that our proposed architecture can achieve faster retrieval time while preserving retrieval accuracy. 

\section{Proposed Architecture}
\label{sec:model}
\subsection{Model Formulation}
Model \textbf{HQ} is shown in Figure~\ref{fig:model}. We use different neural networks to extract continuous features from different content modalities. For the image network, we can train AlexNet~\cite{alexet} or VGG-Net~\cite{vggnet}, and for the text network, we can use the multi-layer perceptron (MLP). Given the continuous features, we learn binary codes and quantization codes by minimizing hash loss and quantization loss simultaneously.

Querying can be done in three steps: (1) compute the Hamming distance between a query item and all items in the database using the binary hash code, (2) select $\alpha N$-out-of $N$ closest items (assume the database size is $N$, $0<\alpha<1$), and (3) compute distances between a query item and the chosen $\alpha N$ items using quantization codes. We use Asymmetric Quantizer Distance (AQD) to measure quantization similarities. This two-stage query is faster than directly computing quantization distances using AQD for all $N$ points, and it is more accurate than only computing Hamming distances using binary codes.

\subsection{Loss Functions}
\label{sec:loss}
There are four parts in the loss function: (1) similarity loss, (2) hashing loss, (3) balance loss and (4) quantization loss. For simplicity, we define $i,j$ as two data points that come from different modalities. They have continuous representation features $f_i$, $f_j$ and the similarity label $s_{ij}$ ($s_{ij}=1$ means $i$ and $j$ are similar, $s_{ij} = 0 $ means they are dissimilar).

\textbf{Similarity loss}
To measure the similarity of features in different modalities, we adopt the Maximum a Posteriori (MAP) estimation. The logarithm MAP is:
\begin{equation}
\log p(f_i,f_j|s_{ij}) \propto \log p(s_{ij}|f_i,f_j)p(f_i)p(f_j) \label{eqn:map}
\end{equation}
In Equation~\ref{eqn:map}, the conditional likelihood for similarity label $s_{ij}$ is:
\begin{equation}
\begin{aligned}
  p(s_{ij}|f_i,f_j) &= \sigma(\langle f_i,f_j\rangle)^{s_{ij}}(1-\sigma(\langle f_i,f_j\rangle))^{1-s_{ij}}
\end{aligned}
\label{eqn:condition}
\end{equation}
where $s_{ij} \in \{0,1\}$, $\sigma(x) = 1/(1+e^{-x})$ is the sigmoid function. Assuming the prior for $f_i$ and $f_j$ is known, the cross entropy loss term is:
\begin{equation}
 L_{sim} = \sum_{{i,j}} (\log(1+e^{\langle f_i,f_j\rangle })-  s_{ij}\langle f_i,f_j\rangle  ). 
\end{equation} 

\textbf{Hash loss} We create a binary code that has the same length as the dimension of the continuous feature. To minimize binary code error, we define the hash loss. Given the binary code $h \in \{-1, +1\}^{n}$, the hash loss is: 
\begin{equation}
L_h =  \sum_{i,j} ( ||f_i-h_i||_F^2+ ||f_j-h_j||_F^2 ).    
\end{equation}

\textbf{Balance loss} To ensure we use the bit information maximally, we follow previous works~\cite{NIPS2008_d58072be,cmfh} and minimize balance loss defined as follows: 
\begin{equation}
   L_b = \sum_{i,j} ( ||f_i\mathbf{1}||_F^2+ ||f_j\mathbf{1}||_F^2 )
\end{equation}
where $\mathbf{1}$ is a vector of 1s.  It balances the number of $+1$ and $-1$ in a generated binary code for each training sample.

\textbf{Quantization loss}
To minimize quantization error, we sum over the results of multiple dictionary books to approximate the original continuous features: $f \approx \sum_{l = 1}^{m} C^lb^l$. Each dictionary book $C^l \in \R^{n \times k}$, binary indicator $b^l \in \{0,1\}^{k}$, and $\norm{b^l}_0 = 1$. Then, we can define the quantization loss as follows:
\begin{equation}
 L_q =  \sum_{i,j} (||f_i-\sum_{l=1}^mC^lb_i^l||_F^2+ ||f_j-\sum_{l=1}^mC^lb_j^l||_F^2)
 \label{eqn:qloss}
\end{equation}
Combining these four loss functions, we obtain the final loss function $L$:
\begin{equation}
    L = \lambda_{sim}L_{sim}+ \lambda_h L_h +\lambda_b L_b + \lambda_q L_q
    \label{eqn:loss}
\end{equation}
where $\lambda_{sim}, \lambda_h, \lambda_b, \lambda_q$ are the hyper-parameters to balance between different losses.
\subsection{Learning Algorithm}
We need to learn neural network parameters $\Theta$, binary code $h$, 
dictionary book $C$ and 1-of-k binary indicator $b$.
$\Theta$
can be optimized by performing back propagation given loss function in Eq.~\ref{eqn:loss}. 
The optimal solution for minimizing $ ||f-h||_F^2$ is $h = sgn(f)$, where $sgn(x) = 1 \text{ if } x \geq 0$, $sgn(x) = -1 \text{ if } x < 0$.
since $h \in \{-1,+1\}$.

We follow~\cite{cdq} when updating $C$ and $b$. $C$ has a closed form solution if we fix other parameters in Equation~\ref{eqn:qloss}: $C = [F_iB_i^T + F_jB_j^T][B_iB_i^T +B_jB_j^T]^{-1}$. $F,B$ are the batched features and indicators for all training samples. $b$ can be updated by thoroughly checking all possible indicators since we restrict $\norm{b}_0 = 1$.  In each epoch, we alternatively update C and b for a fixed number of times until they converge.

\section{Retrieval Efficiency Analysis}
In this section, 
we analyze the computation and memory complexity of HQ and other hashing methods in the cross-modal retrieval task. We show the complexity comparison between the lossless model, the quantization model, and the hashing model in Table~\ref{tab:complexity}. We include detailed explanations of the computation and memory complexity and approximation error analysis in the supplemental material.

\begin{table*}
\centering
    \scalebox{0.75}{%
    \begin{tabular}{|c|c|c|}
    \hline
        \bf Model &  \bf Memory  & \bf{Computation}\\
        \hline
        Lossless &$O(32Nn)$&$O(Nn)$ \\
        \hline
        Quantization & $O(32mkn+Nm\log_2k)$ &$O(mkn+Nm)$\\
        \hline
        Binary hash & $O(Nn)$& $O(Nn)$\\
        \hline 
        HQ  & $O(Nn+32mkn+Nm\log_2k)$& $O(Nn+mkn+\alpha Nm)$\\
        \hline
    \end{tabular}
}
  \caption{Memory and computation analysis for different models. We assume the total number of items in the database is $N$, dense feature dimension $n$, each of the $m$ quantization dictionary book has length $k$. }
    \label{tab:complexity}
\end{table*}

\label{sec:efficiency}
Now assume two models: (1) A model with hashing and quantization as described in Figure~\ref{fig:model-t} (HQ); it consists of  $n$-length binary code and $m_1$ $n \times k_1$ dictionaries. After hamming distance search, it narrows down the database size $N$ to $\alpha N$ (0< $\alpha \ll 1$). (2) A model with only quantization codes; it consists of $m_2$ $n \times k_2$ dictionaries (CDQ~\cite{cdq}).

\noindent\textit{\textbf{Claim.}
If Model one and two use the same amount of memory, retrieval time for Model one will be $O(n)$ times faster, where $n$ is the continuous feature dimension, with some assumptions. }
\begin{proof}
The dominant terms in memory storage for Model one and two are $O(Nn+Nm_1\log_2 k_1)$ and $O(Nm_2\log_2 k_2)$. To make them equal, we have $O(n+m_1\log_2 k_1)  = O(m_2\log_2 k_2)$. We assume $log_2 k_1 \approx p_1k_1$  and $log_2 k_2 \approx p_2k_2$ and $p_1 = p_2= p$, where $p$ is a very small constant number $(0< p \ll 1)$. Therefore, $O(m_2k_2) = O(n/p+m_1k_1)$.

In terms of computation complexity, the dominant terms for Model one and two require $O(Nn+ \alpha Nm_1+m_1nk_1)$ and $O(Nm_2+m_2nk_2)$. The computation complexity of case two over case one is:
\begin{equation}
\frac{O(Nm_2+m_2nk_2)}{O(Nn+ \alpha Nm_1+m_1nk_1)} = \frac{O(Nm_2+m_1nk_1+n^2/p)}{O(\alpha N m_1+m_1nk_1+Nn)}.
\label{eqn:ratio}
\end{equation}
The right side is derived by assuming Model one and two use the same amount of memory, i.e. $O(m_2k_2) = O(n/p+m_1k_1)$.

We can see that the retrieval time advantage is proportional to $\alpha$. The more we narrow down the search space, the faster the retrieval speed is. Also, the advantage is proportional to the feature size $n$: we gain more benefit if feature length is large. If we assume $O(Np) = 1$, Equation~\ref{eqn:ratio} becomes $O(n)$.
\end{proof}

\section{Experiments}

\subsection{Dataset}
\label{sec:dataset}
We used NUS-WIDE dataset~\cite{nuswide}, MIR-Flickr dataset~\cite{huiskes08} and Amazon Review~\cite{amazonreview} for experiments. \textbf{NUS-WIDE} public web image dataset contains images associated with textual tags.  Besides, each image-tag pair is annotated with one or multiple labels from 81 concept labels. 
 Following the prior work~\cite{p1}, we used a subset of 195,834 image-tag pairs that belong to at least one of the 21 most frequent labels.  We randomly sampled 10,500 pairs as training data, 1,050 pairs as validation data, and 2,100 pairs as test data. 
\textbf{MIR-Flickr} consists of 25,000 images collected from the Flickr website. Each image-tag sample belongs to at least one of the 24 labels.
We randomly sampled 5,000 pairs as training data, 1,000 pairs as validation data, and 2,000 pairs as test data. For \textbf{Amazon} dataset, we used image-title pairs from Grocery and Gourmet Food category. Each image-title belongs to at least one of the 14 categories. 
We randomly sampled 10,000 pairs as training data, 320 pairs as validation data, and 2,000 pairs as test data. 

The similarity measurement is based on the concept labels associated with the images and tags. Each pair is similar if they share at least one same label, otherwise they are dissimilar.
We randomly shuffled the tags and re-paired them with images to create dissimilar pairs. After the shuffle process, we have around 31\%, 53\%, and 21\% similar pairs for NUS-WIDE, MIR-Flickr, and Amazon datasets respectively.

\subsection{Implementation Details}
\label{sec:implementation}
To extract image features in NUS-WIDE and MIR-Flickr, we initialized image feature extraction networks with pre-trained VGGNet-19~\cite{vggnet} for Section 5.3 and with AlexNet~\cite{alexet} pre-trained on the ImageNet dataset~\cite{imagenet_cvpr09} for Section  5.4. 
Amazon image features are provided by the original paper~\cite{amazonreview}. In all datasets, final image features are 128-length vectors.
For text features, we followed previous work~\cite{dcmh} and extracted 1,000, 1,386 and 1,000 most frequently used tags/words for NUS-WIDE, MIR-Flickr and Amazon, respectively, and created bag-of-word vectors. These vectors were converted to the final 128-length text features using a two-layer MLP. 

We cross-validated the hyper parameters and finally set  $\lambda_q$ = 0.0001, $\lambda_{sim}$ = 70, $\lambda_h$ =0.01, $\lambda_b$ = 0.01 for NUS-WIDE,  $\lambda_q$ = 0.0001, $\lambda_{sim}$  = 50, $\lambda_h$ = 0.01, $\lambda_b$ = 0.01 for MIR-Flickr and $\lambda_q$ = 0.001, $\lambda_{sim}$  = 10, $\lambda_h$ = 0.01, $\lambda_b$ = 0.001 for Amazon experiments. $\lambda_q$ was set especially small because we observed quantization loss decreasing much faster than other losses. For all experiments, $\alpha N = 100$.  All experiments were performed on a NVIDIA GeForce GTX 1080 Ti. All codes were written with PyTorch 1.1.0. The code is available in Github.~\footnote{https://github.com/rakutentech/hq}.

\subsection{Evaluation Metrics}
We performed two tasks: (1) $I\rightarrow T$: retrieve relevant texts given an image query and (2) $T\rightarrow I$: retrieve relevant images given a text query. We followed prior works~\cite{p2,DBRCH} and used \textit{Mean Average Precision} (MAP)@50 to evaluate the performances.
We also computed the \textit{Harmonic mean} of MAP@50 to measure the balanced performance between both tasks.

\begin{table}[t]
\begin{center}
   \scalebox{0.62}{  
    \begin{tabular}{|c|c|c|c|c|c|c|c|c|c|c|c|c|c|}
       \hline
       \multirow{2}{*}{Task}& \multirow{2}{*}{Method} & \multicolumn{4}{c|}{NUS-WIDE}&  \multicolumn{4}{c|}{MIR-Flickr}&  \multicolumn{4}{c|}{Amazon} \\
       \cline{3-14}
       && 16 &32& 64&128&16 &32& 64&128&16 &32& 64&128  \\

       \hline
         \multirow{9}{*}{$I \rightarrow T$} &
          Lossless &0.521&0.525&0.526&0.528&0.621&0.623&0.623&0.617&0.308 &0.309& 0.345&0.376\\
         \cline{2-14} 
         & CVH~\cite{cvh} &0.372 &0.363 &0.404&0.390&0.606& 0.599&0.596 &0.589& - &- & -&-\\
          &LCMH~\cite{lcmh} &0.354&0.361&0.389&0.383&0.559&0.569&0.585&0.593& - &- & -&-\\
          & QBH~\cite{QBH} &0.524&0.439 &0.266&0.329 & 0.573&0.495&0.401& \bf{ \textcolor{g}{0.602}} & 0.259  & 0.259 &0.260& 0.291 
 \\
          & DJSRH~\cite{djsrh} &0.479& \bf{ \textcolor{g}{0.493}}&0.494 & 0.464&0.582&0.580&0.575& 0.572 & 0.287& 0.300& 0.305& 0.334\\
          & JDSH~\cite{jdsh} &0.370 & 0.371 & 0.372 & 0.370 & 0.558& 0.553& 0.556 &0.561& 0.272& 0.277 & 0.273& 0.280 \\
           \cline{2-14}
            &CMSSH~\cite{cmr-1}& 0.351& 0.353& 0.353&0.370 &0.605&0.602&0.579&0.536 & 0.283 &
0.270 & 0.280 
&0.288 \\
           &DCMH~\cite{dcmh} &0.351&0.467&0.413&0.389&0.628& 0.633&0.609&0.595 &  0.274 & 0.292 &0.294&  0.305 \\
           &CDQ~\cite{cdq} & 0.498& 0.491 & 0.508&  \bf{ \textcolor{g}{0.515}}& \bf{ \textcolor{g}{ 0.638}}& \bf{ \textcolor{g}{0.615}}& 0.604& 0.585 & 0.265&0.279& 0.286& 0.312  \\
         &HQ & \bf{ \textcolor{g}{0.542}} &0.449 & \bf{ \textcolor{g}{0.508}} & 0.411 &0.585&0.583& \bf{ \textcolor{g}{0.610}} &0.583& \bf{ \textcolor{g}{0.293}}  & \bf{ \textcolor{g}{0.301}}&\bf{ \textcolor{g}{0.332}} & \bf{ \textcolor{g}{0.354}}  \\ 
         \hline
         \multirow{9}{*}{$T \rightarrow I$} &
          Lossless &0.489&0.485&0.552&0.548&0.641&0.630&0.648&0.643& 0.309& 0.310& 0.324&  0.354\\
         \cline{2-14} 
         & CVH~\cite{cvh}& 0.401 &0.384& 0.442 & 0.432 & 0.591 &0.583 &0.576 &0.576& - &- & -&-\\
          &LCMH &0.376&0.387&0.408&0.419&0.561&0.569&0.582&0.582& - &- & -&-\\
          & QBH~\cite{QBH} &  0.250&0.294&0.447&0.337& 0.604&0.614&0.523&0.521 &  0.273 & 0.274 & 0.274&0.281 \\
          & DJSRH~\cite{djsrh} &0.365&0.367&0.378&0.384 &0.578&0.589&0.583&0.583&  0.280&  0.275& 0.273& 0.272  \\
          &JDSH~\cite{jdsh} & 0.435 & 0.449 & 0.458& \bf{ \textcolor{g}{0.460}} & 0.616 & 0.598& 0.602&0.612& 0.287 &  0.285& 0.282& 0.302  \\
           \cline{2-14}
            
           & CMSSH~\cite{cmr-1}  &0.393 & 0.393&0.383 &0.381 &0.560 &0.588&0.603 & 0.607 & 0.274& 0.271 & 0.273& 0.304\\
           &DCMH~\cite{dcmh} &0.392&0.358&0.367&0.406&0.619&0.621&\bf{ \textcolor{g}{0.647}}&0.603 & 0.298& 0.284 & 0.290 & 0.289  \\
           &CDQ~\cite{cdq} &0.419 & 0.422 & 0.525&0.433& 0.606 & 0.601& 0.626& 0.623& 0.283 & 0.299& 0.311 & 0.345\\
         &HQ &\bf{ \textcolor{g}{0.493}} &\bf{ \textcolor{g}{0.473}} &  \bf{ \textcolor{g}{0.547}} & 0.432 & \bf{ \textcolor{g}{0.628}}&\bf{\textcolor{g}{0.640}}&0.630& \bf{ \textcolor{g}{0.625}} &\bf{ \textcolor{g}{ 0.305}} &  \bf{ \textcolor{g}{0.309}} &  \bf{ \textcolor{g}{0.322}} &  \bf{ \textcolor{g}{0.332}} \\ 
         \hline
         \hline
        \multirow{9}{*}{Harmonic mean} &
          Lossless &0.504& 0.504& 0.539& 0.538& 0.631 & 0.626 & 0.635 & 0.632& 0.308 & 0.309 & 0.334 & 0.365\\
          \cline{2-14}
        & CVH~\cite{cvh} &0.386 & 0.373 & 0.422&0.410& 0.598&0.591 & 0.586& 0.582&- &- & -&-\\
        & LCMH~\cite{lcmh} &0.365 & 0.374 &0.398 & 0.400 & 0.560&0.569&0.583&0.587&- &- & -&-\\
        &QBH~\cite{QBH} &0.339& 0.352 & 0.334& 0.333& 0.588 & 0.548 & 0.454& 0.559 &0.266& 0.266  &0.267& 0.286\\
        & DJSRH~\cite{djsrh}& 0.414 &0.421 &0.428 &0.420 &0.580 &0.584 &0.579 & 0.577&0.283 & 0.287&  0.288& 0.300\\
        & JDSH~\cite{jdsh} & 0.400& 0.406& 0.411& 0.410 &0.586 &0.575&0.578&0.585 &0.279 & 0.281&  0.277 & 0.291\\
        \cline{2-14}
        & CMSSH~\cite{cmr-1} &0.371 & 0.372 & 0.367 & 0.375 & 0.582 & 0.595 &0.591 & 0.569& 0.278 & 0.270& 0.276& 0.296\\
        & DCMH~\cite{dcmh} &0.370 & 0.405 & 0.389 & 0.397 & \bf{ 0.623} & \bf{ 0.627} & \bf{ 0.627} & 0.599& 0.285 & 0.288& 0.292& 0.297\\
        & CDQ~\cite{cdq} &  0.455& 0.454& 0.516&\bf{0.470}&0.622&0.608&0.615&0.603& 0.274& 0.289& 0.298& 0.328\\
       &  HQ&\bf{ 0.516} &\bf{ 0.461} &\bf{ 0.527} & 0.421 &0.606 & 0.610&0.620& \bf{ 0.603}& \bf{ 0.299}& \bf{ 0.305}& \bf{ 0.327}& \bf{ 0.343}\\
         \hline
    \end{tabular}
    }       
    \end{center}
        \caption{The MAP@50 of our approach and previous models varying compact code dimension. Harmonic mean is based on $I \rightarrow T$ and $T \rightarrow I$ tasks. (CVH and LCMH results are cited from the previous paper~\cite{udcmh}.)}
\label{tab:sota}
\end{table}
\subsection{Results}
\label{sec:differentA}
In this section, we compare our model with 7 previous models: unsupervised cross-modal hashing methods CVH~\cite{cvh}, LCMH~\cite{lcmh}, QBH~\cite{QBH}, DJSRH~\cite{djsrh}, JDSH~\cite{jdsh}, and supervised methods CMSSH~\cite{cmr-1}, CDQ~\cite{cdq} and DCMH~\cite{dcmh}.  We trained and tuned DJSRH and JDSH using the code provided by authors\footnote{https://github.com/zzs1994/DJSRH, https://github.com/KaiserLew/JDSH}. We carefully implemented and tuned QBH, CMSSH, CDQ and DCMH. We also provided Lossless model results as references. Lossless model only uses a similarity loss and returns continuous-valued features. At the retrieval time, it computes cosine similarity between features.
The performance results are listed in Table ~\ref{tab:sota}.\\
\textbf{Comparison to unsupervised models.}
Compared to the unsupervised method DJSRH, HQ achieved a relative improvement of 14.37\% of average Harmonic mean with different code dimensions on NUS-WIDE, 5.05\% on MIR-Flickr, and 9.95\% on Amazon. DJSRH outperformed other unsupervised and some  supervised methods. This is because DJSRH measures similarities in both single and cross modalities with more parameters. For example, DJSRH has almost twice the parameters of HQ\footnote{DJSRH requires 120M parameters while HQ requires 62M. The number of parameters was obtained by PyTorch's model.parameters().} when both models use 64-length codes. 
In addition, we noticed that the MAP improvements on NUS-WIDE and Amazon were higher than the ones on MIR-Flickr when comparing supervised and unsupervised models. This is because NUS-WIDE and Amazon datasets have larger variances in image and text (In NUS-WIDE and Amazon, the average number of labels for one image were 2.2 and 1.5, while it was 3.7 for MIR-Flickr.), which makes retrieval difficult without supervision. \\
\textbf{Comparison to supervised models.}
CMSSH uses a boosting algorithm to learn a linear projection of the original feature and then converts it to binary codes. It does not utilize any deep neural nets. Thus, it is less accurate than the others.
When we compared CDQ (the quantization model) with DCMH (the binary hashing model), CDQ boosted the average Harmonic mean on NUS-WIDE and Amazon by 21.5\% and 3.4\%, while  decreasing 1.1\% on MIR-Flickr.
Compared to the second best model CDQ, HQ achieved an average Harmonic mean improvement of  1.75\% and 7.24\% on NUS-WIDE and Amazon.\\ 
\textbf{Comparison to the model that uses both hashing and quantization.}
 Compared to QBH, which learns hashing and quantization codes sequentially, HQ achieved better performances. There are two possible reasons: (1) QBH cannot fully utilize the quantizer since it is restricted by hash codes; accordingly, the quantization error was calculated based on the assumption that data with the same hash code should share the same dictionary, and (2) QBH uses Hamming distance while HQ uses asymmetric quantizer distance which is more accurate, especially when the hash code length is short. To remove the measurement bias, we evaluate the performance of \textbf{HQ-b} which is HQ using the same distance measurement with QBH: Hamming distance between the query's and database's quantization codes in the second-stage retrieval. In Table~\ref{tab:hq-b}, we observe that with the same distance measurement, HQ-b still outperforms QBH. 

\begin{table}[]
    \centering
    \begin{minipage}{0.48\linewidth}
    \scalebox{0.73}{
\begin{tabular}{|c|c|c|c|c|c|}
       \hline
       \multirow{2}{*}{Task}& \multirow{2}{*}{Method} & \multicolumn{4}{c|}{Amazon} \\
       \cline{3-6}
       && 16 &32& 64&128 \\
       \hline
         \multirow{2}{*}{$I \rightarrow T$} & QBH & 0.259 &0.259&0.260& 0.291\\
         &HQ-b& \bf{ \textcolor{g}{0.286}}&\bf{ \textcolor{g}{0.298}}& \bf{ \textcolor{g}{0.295}}& \bf{ \textcolor{g}{0.310}}  \\
          \hline
         \multirow{2}{*}{$T \rightarrow I$} & 
          QBH & 0.273& 0.274& 0.274& 0.281\\
         &HQ-b&\bf{ \textcolor{g}{ 0.289}}& \bf{ \textcolor{g}{0.290}}& \bf{ \textcolor{g}{0.288}}& \bf{ \textcolor{g}{0.306}}\\
         \hline
          \hline
         \multirow{2}{*}{H-Mean} & QBH & 0.266& 0.266& 0.267& 0.286\\
         &HQ-b&  \bf{0.287} & \bf{0.294}& \bf{0.291} & \bf{0.308}\\
         \hline
    \end{tabular}
    }
    \caption{MAP@50 of HQ-with-Hamming-distance evaluation.}
    \label{tab:hq-b}
   \end{minipage}
   \hfill
   \begin{minipage}{0.48\linewidth}
 \scalebox{0.7}{  
 \begin{tabular}{|c|c|c|c|c|c|}
       \hline
       \multirow{2}{*}{Task}& \multirow{2}{*}{Method} & \multicolumn{4}{c|}{Amazon} \\
       \cline{3-6}
       && 16 &32& 64&128 \\
       \hline
         \multirow{3}{*}{$I \rightarrow T$} & CDQ+L1 & 0.259 & 0.289 & 0.259 & 0.264 \\
         &CDQ+L2 & 0.284&0.292 & 0.282 & 0.318   \\
         &HQ &  \bf{ \textcolor{g}{ 0.293}}  &\bf{ \textcolor{g}{0.301}}&\bf{ \textcolor{g}{0.332}} &\bf{ \textcolor{g}{0.354}} \\
          \hline
         \multirow{3}{*}{$T \rightarrow I$} & 
          CDQ+L1 &  0.296 & 0.292&  0.318 & 0.297\\
         &CDQ+L2 &0.288& 0.293& 0.305  & 0.296\\
         &HQ & \bf{ \textcolor{g}{ 0.305}} & \bf{ \textcolor{g}{0.309}} & \bf{ \textcolor{g}{0.322}} &  \bf{ \textcolor{g}{0.332}} \\
         \hline
          \hline
         \multirow{3}{*}{H-Mean} & CDQ+L1 & 0.276& 0.290& 0.285& 0.280 \\
         &CDQ+L2 &0.286& 0.292& 0.293&0.307 \\
         &HQ & \bf{0.299}& \bf{0.305}& \bf{0.327}& \bf{0.343 } \\
         \hline
    \end{tabular}
   }
     \caption{MAP@50 of HQ and CDQ-with-regularizer models varying compact code dimension.}%
      \label{tab:amazonregularizer}
   \end{minipage}   
\end{table}

\begin{figure}[t]
\centering
\begin{minipage}{.45\linewidth}
    \includegraphics[width=\linewidth]{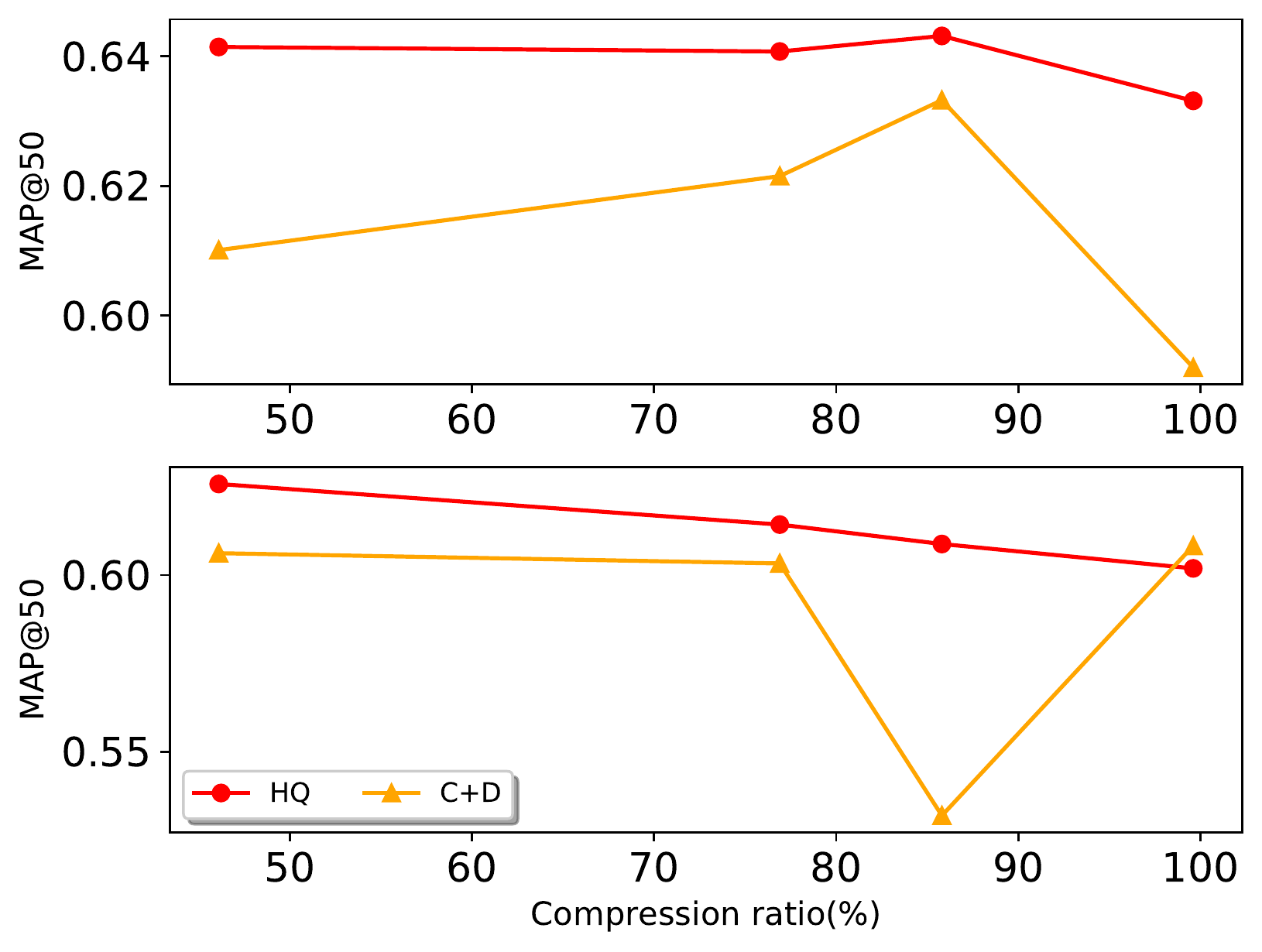}
        \caption{MAP@50 of models with different code-learning strategy on MIR-Flickr. ($I\rightarrow T$ on the top, $T\rightarrow I$ on the bottom)}
    \label{fig:cdcomparison}
\end{minipage}
\hfill
\begin{minipage}{.45\linewidth}
    \includegraphics[width=\linewidth]{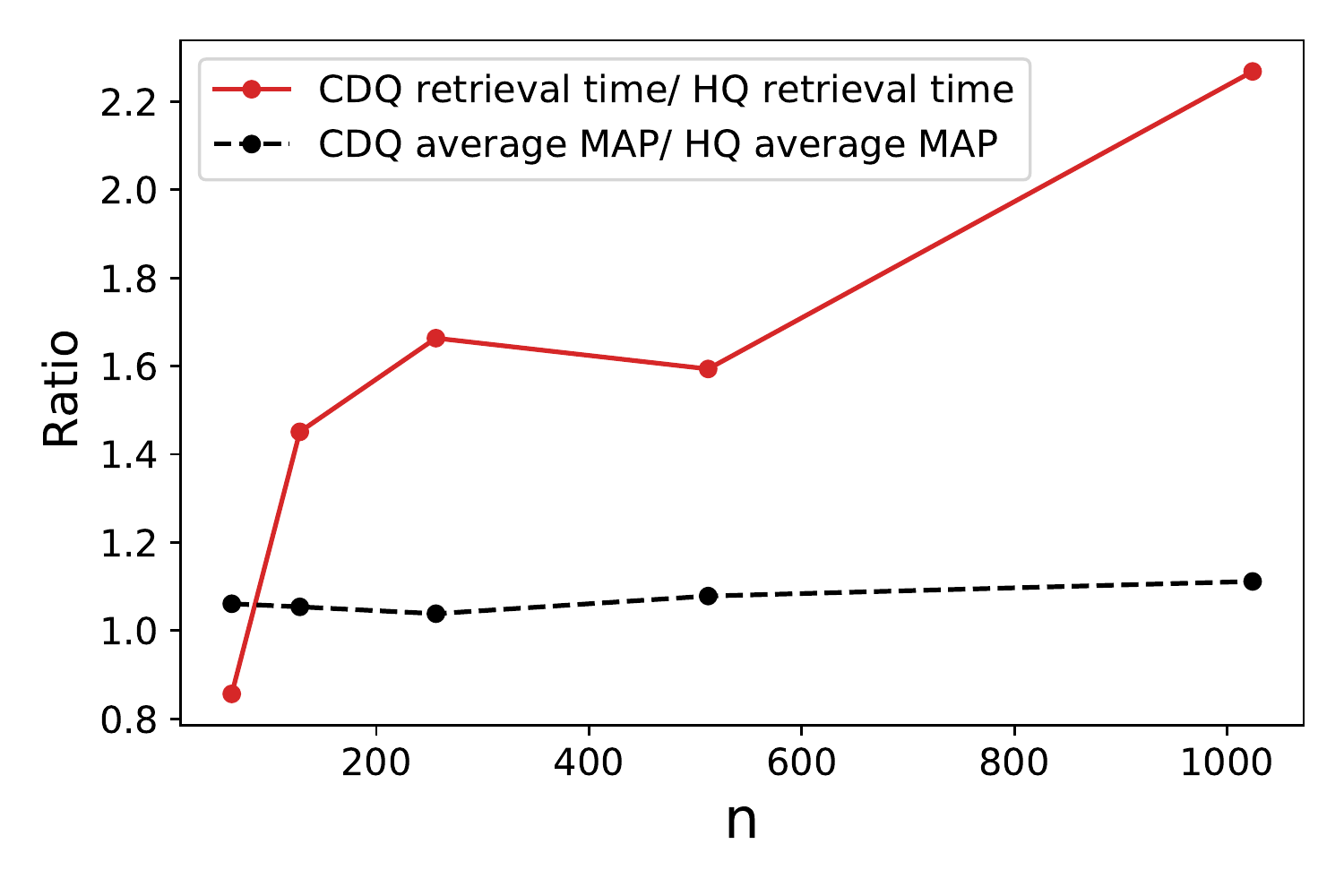}
    \caption{Retrieval efficiency and quality of HQ and CDQ on NUS-WIDE varying $n$. Average MAP is the average of $I\rightarrow T$ and $T\rightarrow I$ MAP@50.}%
  \label{fig:varyn}
\end{minipage}
\end{figure}

\subsection{HQ Effectiveness Analysis} 
\label{sec:results}
In this section, we demonstrate the effectiveness of HQ in terms of learning strategy, retrieval strategy, and  retrieval efficiency.\\
\textbf{Impact of learning two compact codes.}
We assume that learning binary code acts as a regularizer for learning quantization code, and vice versa. To verify this assumption, we implemented CDQ with L1 and L2 regularizers on the feature extraction neural networks. The result is shown in Table~\ref{tab:amazonregularizer}. Indeed, CDQ with L1/L2 regularizer underperformed HQ and thus verified our assumption.\\ 
\textbf{ Impact of simultaneous code-learning.} HQ learns quantization codes and binary codes together. We compare HQ with another model which uses a different learning strategy while keeping the same loss and the retrieval method: \textbf{C+D}.  It trains a CDQ and DCMH  separately, and uses the two-stage retrieval process; it uses DCMH to narrow down search space and  retrieve final results using CDQ. This model learns quantization codes and binary hash codes separately. We show MAP@50 of both models varying compression ratio in Figure~\ref{fig:cdcomparison}. 
The compression ratio is calculated based on the memory required for storing the features. The compression ratio is $1-\frac{P_*}{P}$ where $P_*$ is the size of memory we need for retrieval in the compressed models, and $P$ is the memory needed in the lossless model (See Table~\ref{tab:complexity}). 
From Figure~\ref{fig:cdcomparison}, C+D did not achieve equally good results compared to HQ. 
Thus, learning hash and quantization codes simultaneously achieves a better result than learning them separately.

\begin{figure}[t]
\centering
\begin{minipage}{.4\linewidth}
    \includegraphics[width=\linewidth]{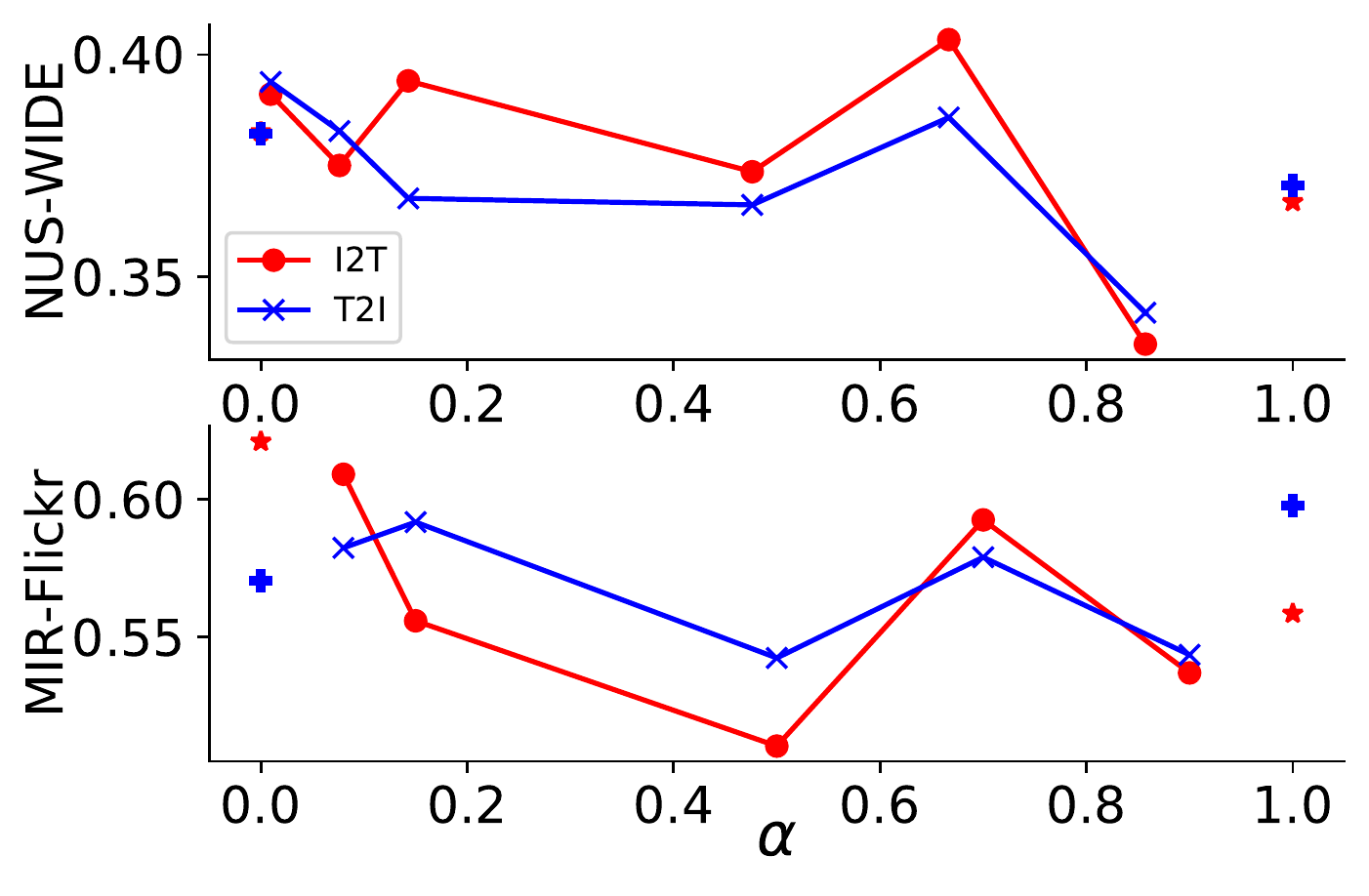}
     \label{fig:experimenti2t}
\end{minipage}
\hfill
\begin{minipage}{.4\linewidth}
    \includegraphics[width=\linewidth]{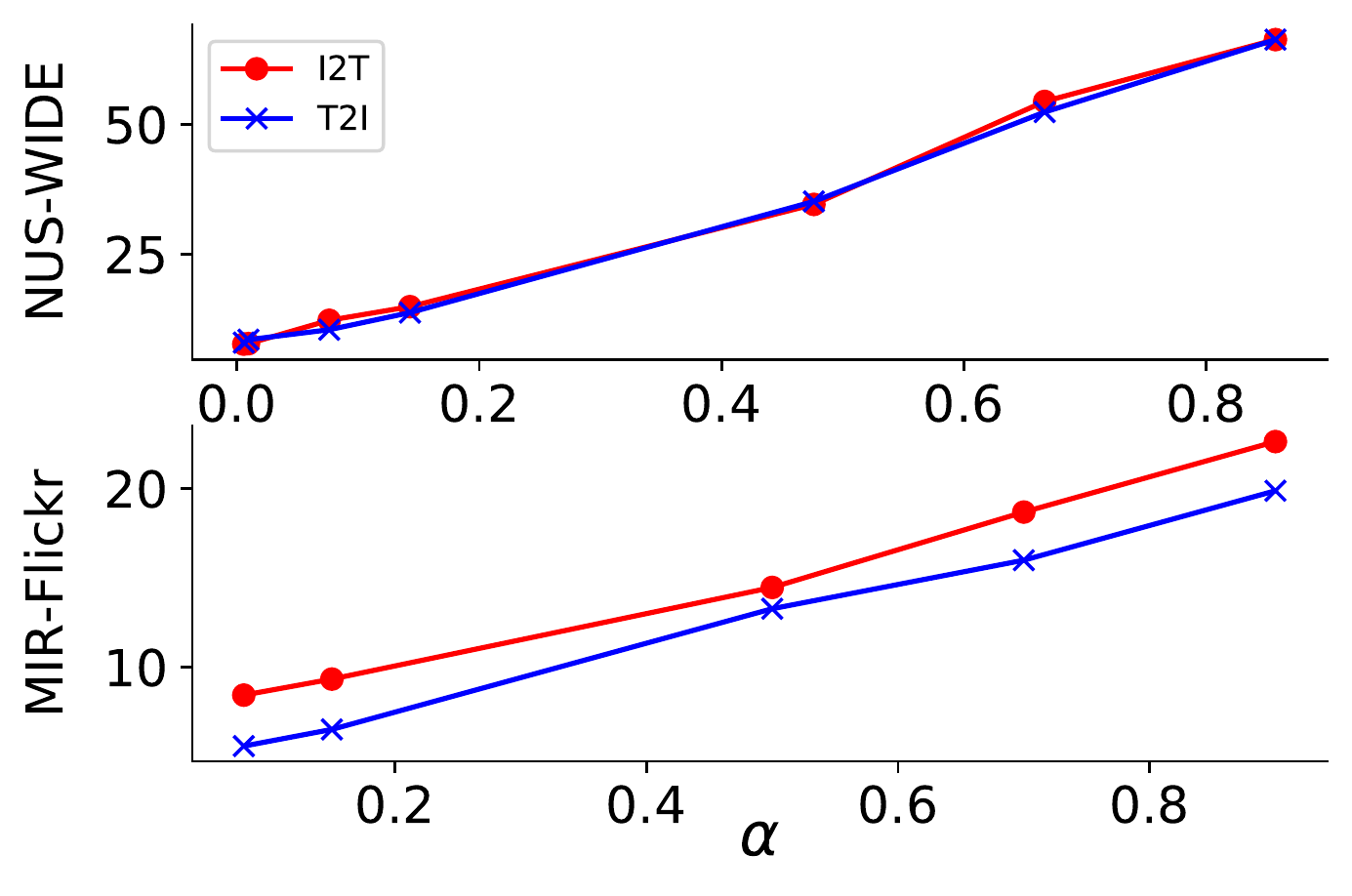}
\label{fig:experimentt2i}
\end{minipage}
    \caption{MAP@50 (left) and retrieval time [$e^{-4}sec$] (right) per query varying $\alpha$.}
    \label{fig:alpha}
\end{figure}

\noindent\textbf{Impact of the two-stage retrieval.} In HQ, we select $\alpha N$-out-of N data points at the first retrieval step. The size of the subset is important as it balances retrieval accuracy and retrieval speed.   
In Figure~\ref{fig:alpha}, we show the MAP@50 and retrieval time while varying $\alpha$. $\alpha=0.0$ indicates that we compute MAP@50 only using binary codes, and $\alpha=1.0$ indicates we only use quantization codes to compute distance scores. We found that (1) the combination of hashing and quantization achieved the best results, (2) the best performance was achieved when $\alpha$ was around $0.1\sim 0.3$ and $0.6\sim0.8$, 
(3) retrieval time was always proportional to $\alpha$. Since we learn both hash and quantization codes simultaneously, the information is diffused to both codes. Therefore, using only quantization for retrieval does not yield best performance.\\
\textbf{Retrieval efficiency}
In Section~\ref{sec:efficiency}, we proved that HQ retrieval process is  $O(n)$ times faster than CDQ, where $n$ is the continuous feature dimension. We computed the retrieval time of the two models varying $n$ while keeping the same memory usage amount in Figure~\ref{fig:varyn}. It showed HQ is almost linearly faster than CDQ while keeping similar retrieval accuracy.

\section{Conclusions}
We proposed HQ, a new cross-modal retrieval structure that combines binary hashing and quantization. It simultaneously learns both codes to preserve semantic information in multiple modalities using an end-to-end deep learning architecture. We also leveraged a two-stage retrieval method for faster and more accurate retrieval results. The experimental results demonstrated that HQ outperformed supervised neural network-based compact coding models.

\clearpage

\bibliography{bmvc_review}

\end{document}


\maketitle
In this supplemental material, we present preliminaries of hashing and quantization in Section~\ref{sec:prelim}. In Section~\ref{sec:analysis}, we show approximation error analysis, and  computation and memory analysis of HQ. Finally, we provide data statistics, implementation details and additional experiments about retrieval efficiency in Section~\ref{sec:exp}.
\section{Preliminaries}
\label{sec:prelim}
\textbf{Binary hashing} Given input $x \in \R^{n}$, we compute binary code $h_x = H(x) \in \R^{n}$, where $H(\bullet)$ is a function that maps continuous value to $\{+1,-1\}$. The similarity measurement is Hamming distance:
\begin{equation}
  \text{dist}(h_x,h_y) = \text{sum } (h_x \text{ XOR } h_y)  
\end{equation}
 The larger the value, the greater the dissimilarity between the two.\\
\textbf{Quantization} Given input $x \in \R^{n}$, we compute the quantization $ x \approx Cb_x $, where $C \in \R^{n \times k}$ is the dictionary book, $b_x \in \R^{k} $ is the index indicator for $x$. We assume the input is assigned to only  one entry of the dictionary: $\norm{b_x}_0 = 1$. Different inputs will share the same $C$ but will have different index indicators.
We use Asymmetric Quantizer Distance (AQD) to measure quantization similarities: 
\begin{equation}
AQD(x,y) = x^T(Cb_y)    
\end{equation} A greater value correlates with a greater similarity between the two.
Generally, quantization preserves more information than binary codes, though it is slower at searching step since AQD requires more computation than Hamming distance.

\begin{figure}[t]
\begin{center}
    \includegraphics[width=2in]{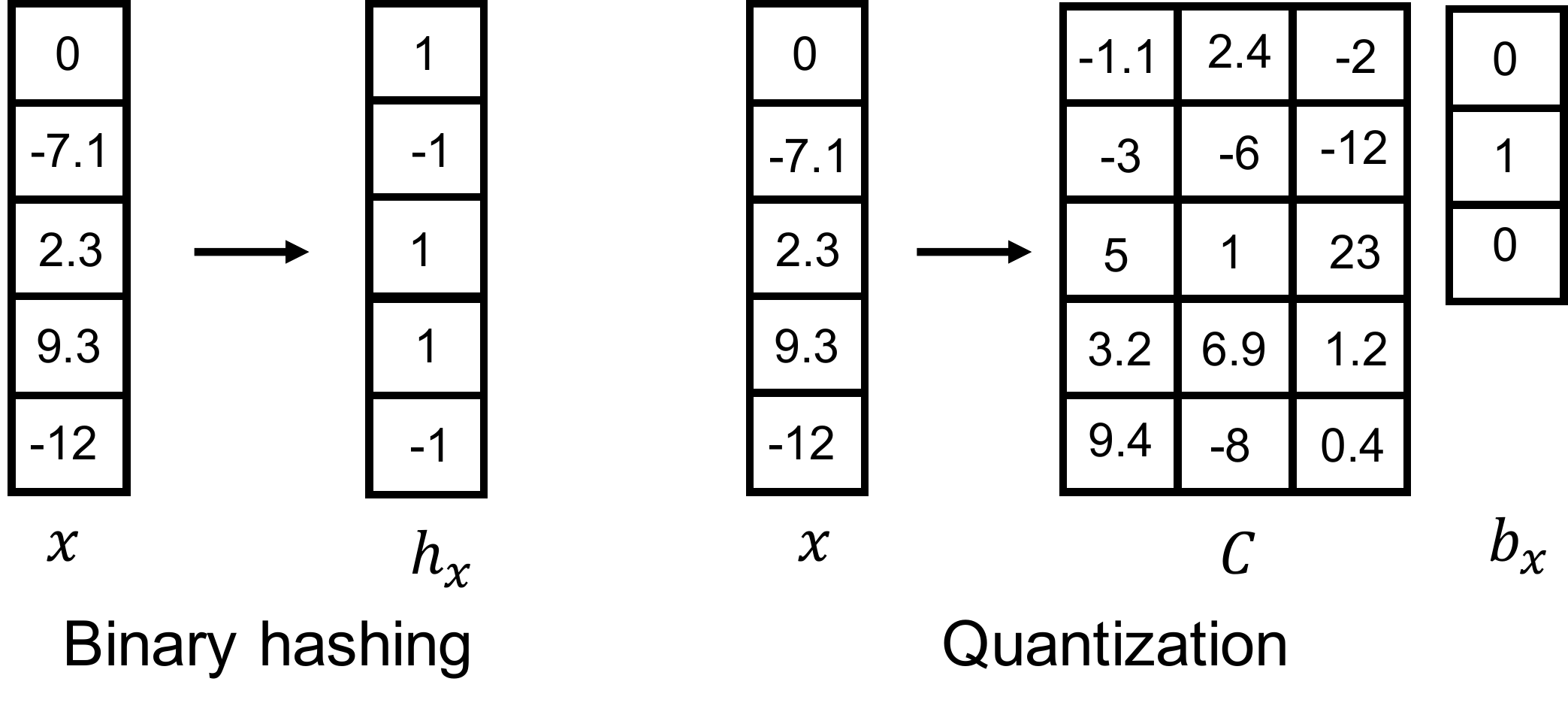}
        \end{center}
    \caption{Binary hashing and quantization.}
\end{figure}

\section{Analysis}
\label{sec:analysis}
In this section, we theoretically show that minimizing HQ's loss function can provide the best compact code error. We also analyze computation and memory complexity of HQ and other hashing methods in the cross-modal retrieval task.

\subsection{Approximation Error Analysis}
We use $h_i, h_j, Cb_i, Cb_j$ and $f_i, f_j$  to represent binary codes, quantization codes and dense features for data $i$ and data $j$ from different modalities.

\textbf{Hashing error}
Even though Hamming distance is used for computing binary code distance, we use the relaxed Euclidean distance to compute the error bound since it is convex.
\begin{equation}
\begin{aligned}
&|d(h_i, h_j) -d(f_i,f_j)| \\
&= | \norm{h_i-h_j}_F^2 - \norm{f_i-f_j}_F^2|   \\
& = | \norm{sgn(f_i)-sgn(f_j)}_F^2 - \norm{f_i-f_j}_F^2|
\end{aligned}
\label{eqn:binary-error}
\end{equation}
Minimizing Equation~\ref{eqn:binary-error} is equal to minimizing the distance between $sgn(f_i)$ and $f_i$. In other words, the continuous feature should be close to +1 or -1. 
We achieved this by adding a \textit{tanh} layer after the final feature layer.

\textbf{Quantization error}
Since AQD approximates inner product distance, we compare AQD with direct inner product results using continuous features $f$.
\begin{equation}
\begin{aligned}
|AQD(i, j) -d(f_i,f_j)|  &= | (f_i)^TCb_j -(f_i)^Tf_j|  \leq \norm{f_i} \norm{f_j-Cb_j}
\end{aligned}
\label{eqn:quantization-error}
\end{equation}
Because $\norm{f_i}$ is a fixed scalar which does not effect the relative distance measurement, we only need to minimize $\norm{f_j-Cb_j}$: the distance between dense feature and quantization feature. This term already exists in the loss function $L_q$. Thus, we are guaranteed to get best quantization error by minimizing our proposed loss function. 

\begin{table*}
\begin{center}
    \scalebox{0.85}{%
    \begin{tabular}{|c|c|c|}
    \hline
        \bf Model &  \bf Memory  & \bf{Computation}\\
        \hline
        Lossless &$O(32Nn)$&$O(Nn)$ \\
        \hline
        Quantization & $O(32mkn+Nm\log_2k)$ &$O(mkn+Nm)$\\
        \hline
        Binary hash & $O(Nn)$& $O(Nn)$\\
        \hline 
        HQ  & $O(Nn+32mkn+Nm\log_2k)$& $O(Nn+mkn+\alpha Nm)$\\
        \hline
    \end{tabular}
}
\end{center}
  \caption{Memory and computation analysis for different models. We assume the total number of items in database is $N$, dense feature dimension $n$, each of the $m$ quantization dictionary book has length $k$ }
    \label{tab:complexity}
\end{table*}
\subsection{Computation and Memory Analysis}
\label{sec:cma}
In this section, we compare the computation and memory efficiency of HQ with the method which only uses quantization.

Memory requirement for quantization is $\bm{O(32mkn+Nm log_2 k)}$: since we have $m$ dictionary books and each book has $k$ $n$-length features, we need $O(32mkn)$ to store the dictionary books when data is in single-precision float(32bits). For each data point, we will use $O(m \log_2 k)$ to store the $m$ one-of-$k$ indicator ($\log_2 k$ bits) $b$. We assume we have $N$ data points.
Memory requirement for binary hashing feature is $\bm{O(Nn)}$. Note that if we use a lossless model which uses continuous features), it will be $\bm{O(32Nn)}$.

For the quantization model, we can pre-compute the dot product between the query and each dictionary with $O(mnk)$ operations and store it in a temporary look-up table, then we need $O(m)$ additions to compute the AQD distance between query and each data point with the look-up table. Overall, we need $\bm{O(mnk+Nm)}$ operations to complete one query retrieval with the quantization model, while we need $\bm{O(Nn)}$ for the hashing model.
We show the complexity comparison between the lossless model, the quantization model,  and the hashing model in Table~\ref{tab:complexity}. Quantization and hashing are more memory efficient than the lossless model.

\begin{figure*}
\begin{minipage}{.45\textwidth}
   \begin{center}
\includegraphics[width=0.95\linewidth]{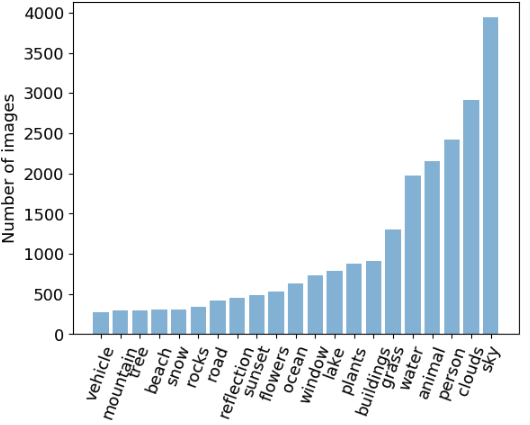}
 \end{center}
  \captionof{figure}{NUS-WIDE Image-label distributions.}
\end{minipage}\hfill
\begin{minipage}{.45\textwidth}
    \begin{center}
  \includegraphics[width=0.95\linewidth]{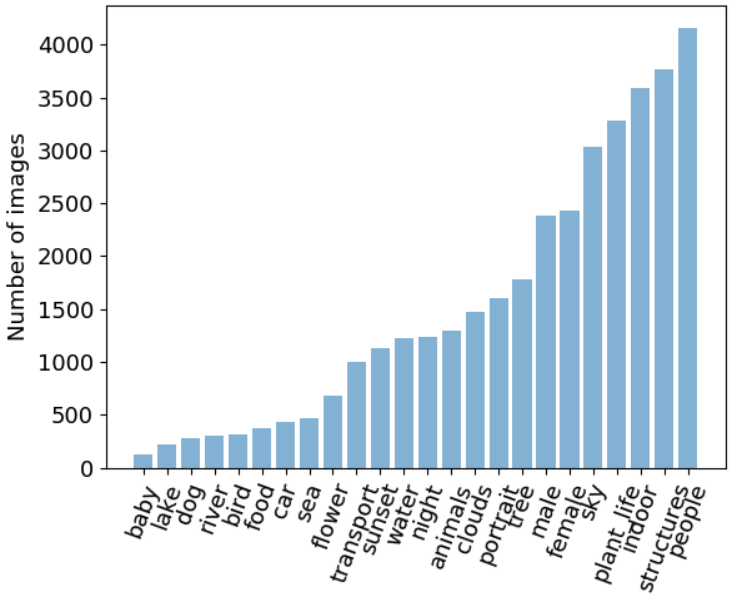}
   \end{center}
  \captionof{figure}{MIR-Flickr Image-label distributions.}
\end{minipage}
\label{fig:stats}
\end{figure*}

\begin{figure}
\begin{minipage}{.45\textwidth}
  \begin{center}
  \includegraphics[width=0.92\linewidth]{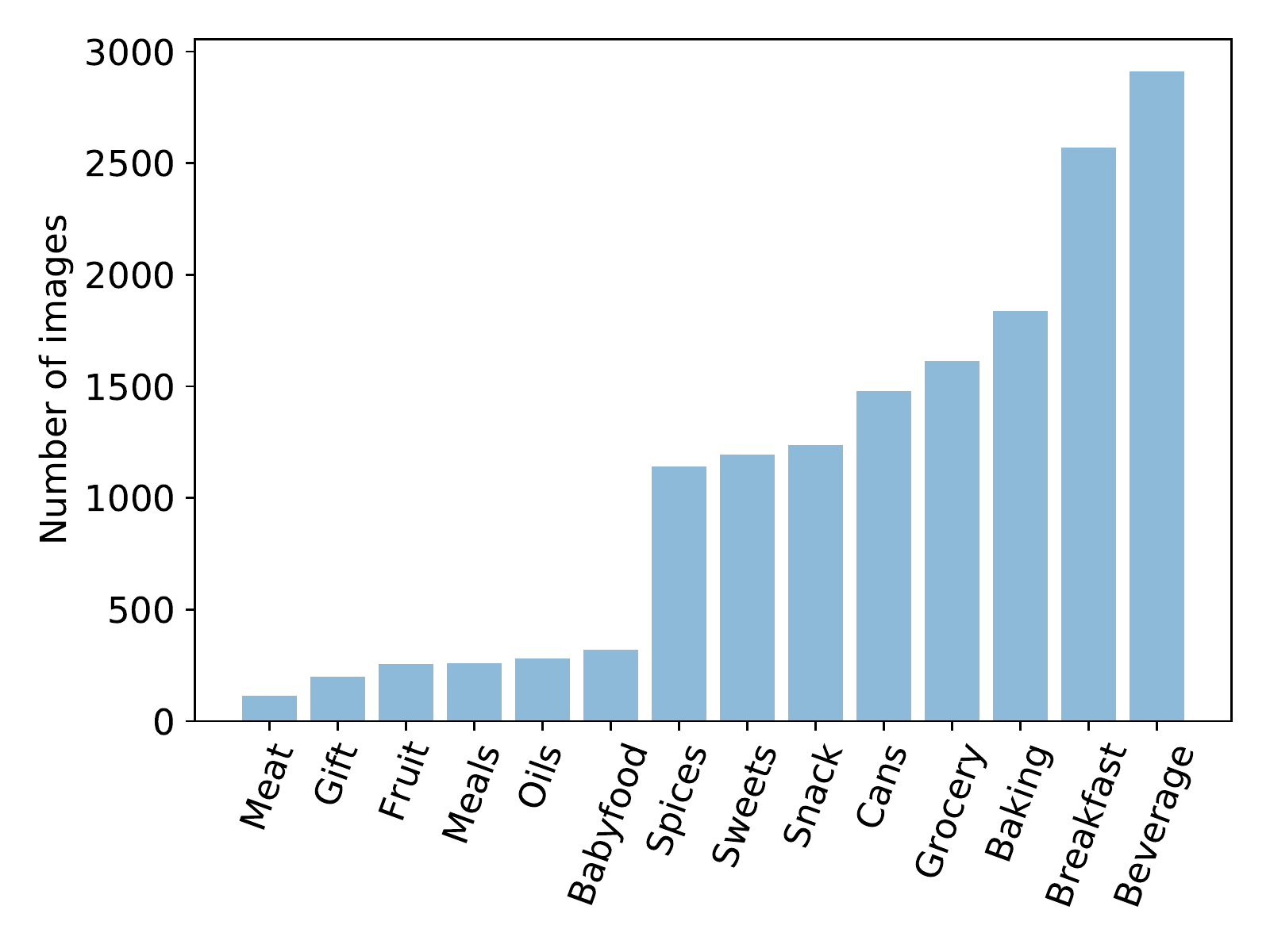}
  \end{center}
  \caption{Amazon Image-label distributions.}
\label{fig:stats-ama}

\end{minipage}\hfill
\begin{minipage}{.45\textwidth}
       \begin{center}
     \includegraphics[width=0.95\textwidth]{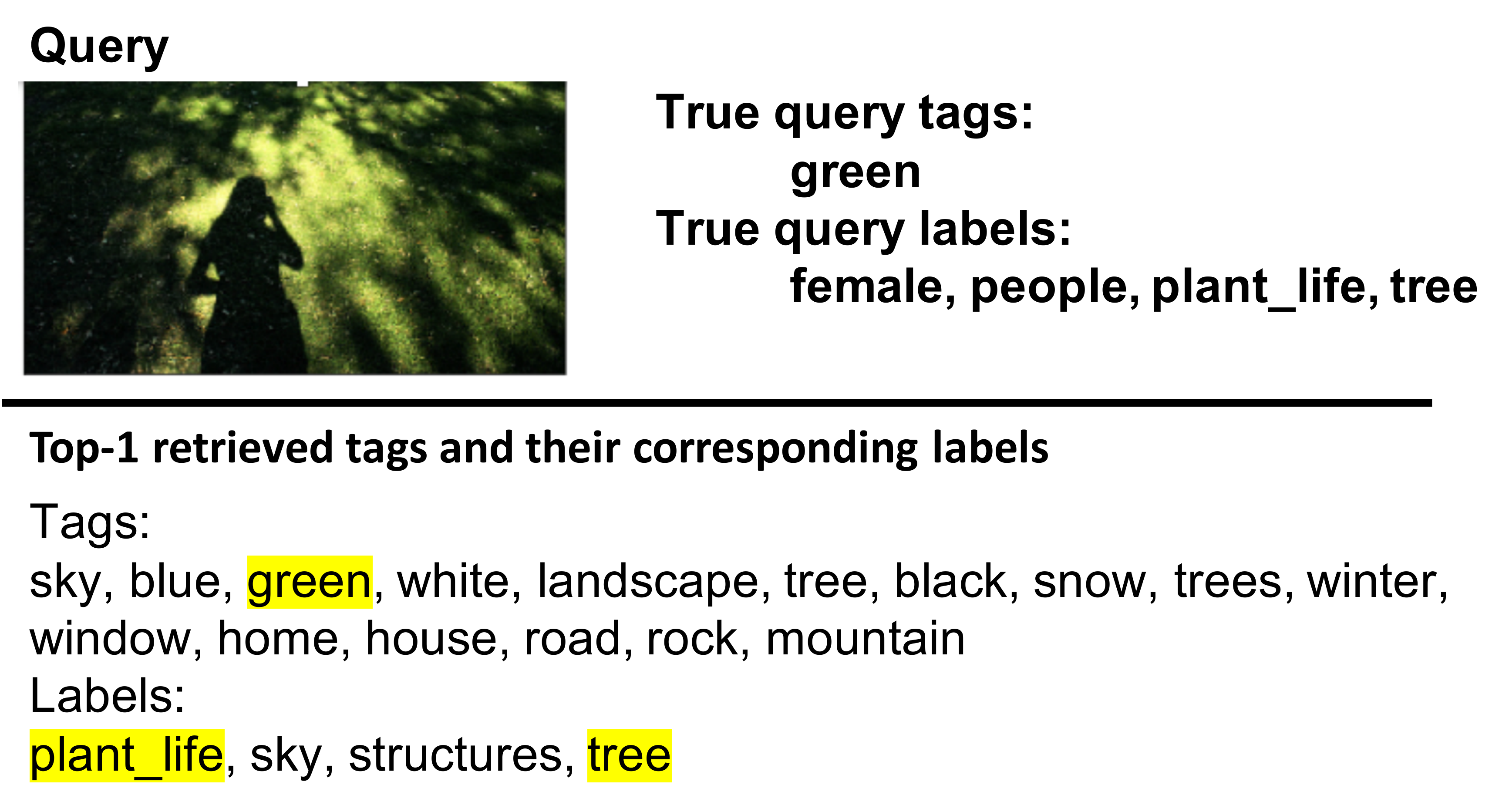}
      \end{center}
     \caption{Retrieval example in MIR-Flickr.}
     \label{fig:example}
     \end{minipage}
 \end{figure}

\section{Experiments}
\label{sec:exp}
\subsection{Dataset}
\label{sec:dataset}
We used NUS-WIDE dataset~\cite{nuswide}, MIR-Flickr dataset~\cite{huiskes08} and Amazon Review~\cite{amazonreview} for experiments. The NUS-WIDE public web image dataset contains images associated with textual tags.  Besides, each image-tag pair is annotated with one or multiple labels from 81 concept labels, such as ``nature,'' ``water,'' and ``sunset.''   
 MIR-Flickr consists of 25,000 images collected from the Flickr website. Each image-tag sample belongs to at least one of the 24 labels, such as ``sky,'' ``car,'' ``lake.''
We selected Amazon Grocery and Gourmet Food category and used their image-title pairs as training/test data. Each image-title belongs to at least one of the 14 categories, such as ``meat,'' ``fruit,'' and ``spices''. 
 
Figure~\ref{fig:example} shows an image-to-text retrieval example. Here, we retrieved a group of text tags in the database for a given query image. From the retrieved tags, we list up their associated labels. Since the highlighted labels exist in the image's labels (i.e. female, people, plant-life, tree), we consider the retrieval is correct.

\subsection{Implementation Details}
\label{sec:implementation}
To extract image features in NUS-WIDE and MIR-Flickr, we initialized image feature extraction networks with pre-trained VGGNet-19~\cite{vggnet} for Section 5.3 and with AlexNet~\cite{alexet} pre-trained on the ImageNet dataset~\cite{imagenet_cvpr09} for Section  5.4. 
Amazon image features are provided by the original paper~\cite{amazonreview}. In all datasets, final image features are 128-length vectors.
For text features, we followed previous work~\cite{dcmh} and extracted 1,000, 1,386 and 1,000 most frequently used tags/words for NUS-WIDE, MIR-Flickr and Amazon, respectively, and created bag-of-word vectors for text inputs. These vectors were converted to the final 128-length text features using a two-layer MLP.

We cross-validated the hyper parameters and finally set  $\lambda_q$ = 0.0001, $\lambda_{sim}$ = 70, $\lambda_h$ =0.01, $\lambda_b$ = 0.01 for NUS-WIDE experiments,  $\lambda_q$ = 0.0001, $\lambda_{sim}$  = 50, $\lambda_h$ = 0.01, $\lambda_b$ = 0.01 for MIR-Flickr experiments and $\lambda_q$ = 0.001, $\lambda_{sim}$  = 10, $\lambda_h$ = 0.01, $\lambda_b$ = 0.001 for Amazon experiments. $\lambda_q$ was set especially small because we observed quantization loss decreasing much faster than other losses. $\alpha N = 100$ is set for all experiments.  All experiments were performed on a NVIDIA GeForce GTX 1080 Ti. All codes were written with PyTorch 1.1.0. 

\subsection{Additional experiments about retrieval efficiency}
Table~\ref{tab:sota} shows the number of model parameters and retrieval time of DJSRH, JDSH, and HQ on NUS-WIDE dataset. Note that JDSH requires less parameters than DJSRH, for it is a simplified DJSRH. While HQ and JDSH have the similar number of parameters, HQ's retrieval time is faster. This proves HQ's efficiency. 
\begin{table}
    \begin{center}
    \begin{tabular}{|c|c|c|c|}
    \hline
    \multirow{2}{*}{Models} &\multirow{2}{*}{$\#$Params. }  &  \multicolumn{2}{c|}{Time} \\
    \cline{3-4}
       & & $ I \rightarrow T$ & $T \rightarrow I$  \\
        \hline
        DJSRH & 120M &   3.19&2.69 \\
        \hline
JDSH& 62M&   2.73& 3.03  \\
\hline
            HQ  & 62M  & 2.62& 2.36 \\
        
        \hline
    \end{tabular}
   \end{center}
\caption{The retrieval time (seconds) on NUSWIDE (a total of 2,100 queries over 13,650 data points, Number of parameters is computed using model.parameters() in Pytorch. All models use 128-dim hash code).}
    \label{tab:sota}
\end{table}

\bibliography{egbib}